\documentclass[useAMS,usenatbib]{mn2e}
\usepackage{graphicx}
\usepackage{txfonts}

\title[Asteroseismology of HS~0507+0434B]{Asteroseismology of the ZZ Ceti star HS~0507+0434B\footnotemark[1]\thanks{Based on data obtained
at the Xinglong station of National Astronomical Observatories, China,
the Lijiang station of Yunnan Astronomical Observatory, China,
the San Pedro M\'artir Observatory, Mexico,
the Bohyunsan Optical Astronomy Observatory, South-Korea,
and the Piszk\'estet\H{o} Observatory, Hungary.}}
\author[J.-N. Fu et al.]{J.-N. Fu$^{1}$\thanks{E-mail:
jnfu@bnu.edu.cn};
N. Dolez$^{2,3}$; G. Vauclair$^{2,3}$; L. Fox Machado$^{4}$; S.-L. Kim$^{5}$; C. Li$^{1}$;
\newauthor
L. Chen$^{1}$; M. Alvarez$^{4}$; J. Su$^{6}$; S. Charpinet$^{2,3}$; M. Chevreton$^{7}$; R. Michel$^{4}$; 
 \newauthor
X.H. Yang$^{1}$; Y. Li$^{6}$; Y.P. Zhang$^{1}$; L. Molnar$^{8,9}$ and E. Plachy$^{8,9}$\\
$^{1}$Department of Astronomy, Beijing Normal University, Beijing, China\\
$^{2}$Universit\'e de Toulouse,UPS-OMP, IRAP, Toulouse, France\\
$^{3}$CNRS, IRAP, 14 avenue E. Belin, 31400, Toulouse, France\\
$^{4}$Observatorio Astron\'{o}mico Nacional, Instituto de Astronom\'{\i}a, Universidad Nacional Aut\'{o}noma de M\'{e}xico, Ensenada, B.C., M\'{e}xico\\
$^{5}$ Korea Astronomy and Space Science Institute, Daejon, Korea\\
$^{6}$Yunnan Astronomical Observatory, Chinese Academy of Sciences, Kunming, China\\
$^{7}$LESIA, Observatoire de Paris-Meudon, Meudon, France\\
$^{8}$Konkoly Observatory, MTA CSFK, H-1121, Konkoly-Thege u 15-17, Budapest, Hungary\\
$^{9}$E\"otv\"os University, H-1117 Budapest, P\'azm\'any P\'eter s\'et\'any 
1/a, Hungary}
\begin{document}

\date{Accepted;  Received}

\pagerange{\pageref{firstpage}--\pageref{lastpage}} \pubyear{2011}

\maketitle

\label{firstpage}

\begin{abstract}
The pulsating DA white dwarfs (ZZ Ceti stars) are $g$-mode non-radial
pulsators. Asteroseismology provides strong constraints on their global
parameters and internal structure. Since all the DA white dwarfs falling
in the ZZ Ceti instability strip do pulsate, the internal
structure derived from asteroseismology brings knowledge for the DA white
dwarfs as a whole group.
HS~0507+0434B is one of the ZZ Ceti stars which lies
approximately in the middle of the instability strip for which we have
undertaken a detailed asteroseismological study.
We carried out multisite observation campaigns in 2007 and from December
2009 to January 2010. In total, 206 hours of photometric time-series have
been collected. They have been analysed
by means of Fourier analysis and simultaneous multi-frequency sine-wave
fitting.
In total, 39 frequency values are resolved including 6 triplets and a number of linear combinations. 
We identify the triplets as $\ell$=1 $g$-modes split by rotation. We derived the period spacing, the rotational splitting and
the rotation rate. From the comparison of the observed periods with the theoretical periods of a series of models we estimate the fundamental parameters
of the star: its total mass  M$_{*}$/M$_{\odot}$ = 0.675, its luminosity
L/L$_{\odot}$=3.5$\times 10^{-3}$, and its  hydrogen mass fraction M$_{H}$/M$_{*}$=
10$^{-8.5}$.
\end{abstract}

\begin{keywords}
stars:white dwarfs -- stars:oscillations -- stars:individual:HS~0507+0434B.
\end{keywords}

\section{Introduction}
As the end products of about 97\% of the stars of the Galaxy,
 white dwarf stars
 offer important clues about their prior evolutionary history.
They also provide potentially the age of the galactic disk and of the globular clusters
they belong to (Winget et al. 1987; Harris et al. 2006) through the age that can be estimated 
from cooling sequences (Ruiz \& Bergeron 2001).

However, determining the age of a white dwarf on its cooling sequence requires knowing its fundamental parameters:
the total mass, the effective temperature, the luminosity, the fractional mass of the hydrogen and/or the helium outer layers, the core composition,etc. 
Asteroseismology of white dwarfs provides a unique tool to explore their internal structure and 
determine those fundamental parameters. The method has been successfully applied to the pulsating pre-white dwarf stars of PG~1159 type,
 the GW Vir stars, e.g. the prototype of the group  
 PG~1159-035 (Winget et al. 1991), the hottest one RXJ~2117+3412 (Vauclair et al. 2002) and the coolest one PG~0122+200 (Fu et al. 2007, 
C\'{o}rsico et al. 2007), 
to the DB pulsators, e.g. GD~358 (Winget et al. 1994, Provencal et al. 2009), PG~1351+489 (Redaelli et al. 2011), and to the DAV stars, either 
for individual pulsator, e.g. HL~Tau~76 (Dolez et al. 2006) or for global study of the group properties (Castanheira \& Kepler 2009; Romero et al. 2012).

Since $\approx$ 80\% of the white dwarf stars are of DA type, the uncertainties on their fundamental parameters have a strong impact on the derived age estimates.
This justifies the effort in determining precise fundamental parameters of DA white dwarfs using asteroseimology of the DA pulsators, the DAV or ZZ Ceti stars.
There are  presently 148 pulsating DA white dwarfs known (Castanheira et al. 2010a, 2010b).
 These stars define a
narrow instability strip in the H-R diagram (or in the log$g$-T$_{eff}$
diagram).
This instability strip is a ``pure"
instability strip, which means that all the DA white dwarfs falling in this
domain of the log$g$-T$_{eff}$ diagram do pulsate (Gianninas et al. 2011). This is an indication that the
internal structure of the ZZ Ceti white dwarfs as derived from asteroseismology
is representative of the DA white dwarfs as a whole group. 

White dwarf stars being the oldest
stars in a given stellar population, they can be used to estimate the age of
the population they belong to. However, to achieve such a goal one must rely
on realistic models of white dwarf stars. The major uncertainties in building
such realistic models come from the uncertainty on their total mass and on
their hydrogen mass fraction as well as from the approximate treatment of
the convection by the mixing length theory parameterized by a mixing
length $\alpha$. The asteroseismology is the only method able to determine
the value of the hydrogen mass fraction and to give accurate total mass
estimates of the ZZ Ceti white dwarfs. In addition, new theoretical
developments on the interaction of the pulsations with convection have
opened a way to constrain the efficiency of the convection in white dwarf stars 
(Wu 2001, Montgomery 2005, Montgomery et al. 2010).

The difficulty in determining the fundamental parameters of the ZZ Ceti stars
from asteroseismology comes from two main sources: 1) the ZZ Ceti stars show
generally few modes simultaneously, in contrast with the theoretical
calculations which predict much more unstable modes than observed, and 2) the
pulsation amplitudes become increasingly variable as the ZZ Cetis evolve towards 
the red edge of the instability strip. Both effects make difficult to
find enough appropriate modes to use the method based on the period spacing which
needs a large enough number of pulsation modes to be observed and identified.
Only ZZ Ceti stars close to the blue edge of the instability strip show constant
pulsation amplitude. But in this case, very few modes are unstable since
the stars are just entering the instability strip. In those stars, the
$\kappa$-mechanism due to hydrogen partial ionization is responsible for
the instability since the fraction of the flux conveyed by convection
is negligible (Dolez \& Vauclair 1981, Winget et al. 1982). As the white dwarfs evolve along
their cooling sequence, the fraction of the flux conveyed by convection
increases. The light curves become more complex as a result of the interaction of the convection with the pulsations. They
show signatures of
nonlinear effects like non sinusoidal pulse shapes which reflect into
linear combinations of ``real" frequencies in the Fourier spectrum.
Modeling precisely the light curve resulting from these inteactions can be used 
to constrain the efficiency of convection (Montgomery 2005, Montgomery et al. 2010).
The asteroseismological analysis of various ZZ Ceti stars through the instability strip 
allows to map the convection efficiency. 

 The DA white dwarf 
HS~0507+0434B is one of those ZZ Ceti star of particular interest which we
intend to study in more details.
Its parameters as derived from earlier spectroscopy give an effective temperature of 11630$\pm 200$ K and log$g$= 8.17$\pm 0.05$
(Fontaine et al. 2003; Bergeron et al. 2004) which placed HS~0507+0434B  approximately in the middle of the instability strip
of the ZZ Ceti pulsators. 
 Fontaine et al. (2003) provide a mass value
of 0.71~$M_{\odot}$ derived from the models of Wood (1995) for carbon
core compositions, helium layers of $M_{\rm He}=10^{-2}M_{\star}$, and hydrogen
layers of $M_{\rm H}=10^{-4}M_{\star}$, and absolute magnitude in $V$ of
$M_V=11\fm99$.
 More recent high signal/noise ratio spectroscopy and atmospheric analysis by Gianninas et al. (2011) 
have shifted the ZZ Ceti instability strip to higher effective temperature. With an  
effective temperature and a surface gravity of 12290 $\pm 186$ K and  log$g$= 8.24$\pm 0.05$ respectively,
HS 0507+0434B still lies close to the middle of the new ZZ Ceti instability strip. 
It is of particular interest because it forms a common proper
motion pair with HS~0507+0434A, which is a 20000 K DA white dwarf
(Jordan et al. 1998).
 Gianninas et al. (2011) give also a higher effective temperature of 21550 $\pm 318$ K for the A component.
Since both members of the pair must have been
formed at the same time, this provides one additional constraint on the modeling
of their evolution. 

HS~0507+0434B was discovered to be a ZZ Ceti 
variable by Jordan et al. (1998). It has been observed subsequently from
single sites only (Kotak et al. 2002; Handler et al. 2002). Those
observations allowed the detection of 10 independent frequencies identified as
$l$=1 gravity modes and of 38 linear combinations of these frequencies (Handler et al. 2002).
The linear
combinations of frequencies were interpreted as resulting from the nonlinear
interaction of the pulsations with convection. The 10 independent frequencies are 
formed of
three triplets (i.e. $l$=1 modes split by rotation) plus one single mode.
This was clearly insufficient to constrain the star internal structure.
To improve this situation, we carried out multisite observation
campaigns in 2007, December 2009 and January 2010
to study HS~0507+0434B.

The goal of this paper is to improve the determination of the fundamental parameters of HS~0507+0434B
using the new data collected during the campaigns of 2007, 2009 and 2010. It is organized as follows: the observations
are described in $\S 2$. $\S 3$ analyses the derived amplitude spectrum. A preliminary asteroseismology of  HS~0507+0434B
is presented in $\S 4$ where are discussed successively the frequency identification and their linear combinations, in 4.1, 
the period distribution and the period spacing, in 4.2, the rotational splitting and the rate of rotation, in 4.3,
 the inclination of the rotation axis, in 4.4, and the mode trapping in 4.5. $\S 5$ discusses the amplitude variations.
From the improved list of pulsation modes resulting from this work, we attempt to constrain the fundamental parameters of 
 HS~0507+0434B by building a set of models which are described in $\S 6$, from which we derived one ``best fit'' model. 
We summarize our results in the conclusions $\S 7$.

\section{Observations and data reduction}

HS~0507+0434B was observed during a one week multisite campaign in December
2007, involving the 2.16-m telescope of the National Astronomical Observatories
of China (NAOC) in Xinglong (XL), the 1.8-m telescope of Bohyunsan Optical
Astronomy Observatory (BOAO) of Korea and the 1-m telescope of Piszk\'estet\H{o}
Observatory (PO) of Hungary. Later on, a four-week multisite run was carried
out from December 2009 to January 2010 with the 2.16-m
telescope in XL, the 2.4-m telescope of Yunnan Astronomical Observatory of
China in Lijiang (LJ) and the 1.5-m telescope of San Pedro M\'artir Observatory
of Mexico (SPM). Tables 1 and 2 list the journals of the observations.

In 2007, an observing cycle of 30 seconds was used with the CCD cameras in
BOAO and PO, while a three-channel photoelectric photometer (PMT) was used to
observe a comparison star, the sky background, and the target simultaneously
with the exposure time of 1 second in XL. These PMT data points
were then summed to a similar sampling rate of 30 seconds. In 2009 and 2010,
CCD cameras were used for all runs with an observing cycle of approximately
30 seconds. 46.1 hours of data were collected in December of 2007
with a duty cycle of 31\%, which leads to the frequency resolution of
1.9 $\mu$Hz. In 2009-2010 we obtained 158.9 hours of data with a duty cycle of 
13.4\%  and a better frequency
resolution of 0.23 $\mu$Hz.

\begin{table}
 \centering
  \caption{Journal of observations in December of 2007. Filter symbols: $B$ = Johnson $B$, $W$ = white light.}
  \begin{tabular}{@{}ccccc@{}}
  \hline\noalign{\smallskip}
Date & Telescope & Filter & Detector & Hours \\
  \hline\noalign{\smallskip}
7 & XL 2.16-m  & $W$ & PMT & 7.03 \\
~ & PO 1-m & $B$ & CCD &  0.64 \\ 
8 & XL 2.16-m & $W$ & PMT & 5.27 \\
~ & BOAO 1.8-m & $B$ & CCD & 7.86 \\
9 & BOAO 1.8-m & $B$ & CCD & 7.53 \\
11 & XL 2.16-m& $W$ & PMT & 9.13\\
~& BOAO 1.8-m & $B$ & CCD & 3.43\\
12 & XL 2.16-m & $W$ & PMT & 8.00\\
~& PO 1-m & $B$ & CCD & 0.68\\
13 & XL 2.16-m & $W$ & PMT & 5.20\\
\noalign{\smallskip}\hline
\end{tabular}
\end{table}

\begin{table}
  \caption{Journal of observations from December of 2009 to January of 2010. Filter: Johnson $B$.}
  \begin{center}\begin{tabular}{ccc}
  \hline\noalign{\smallskip}
Date & Telescope & Hours \\
  \hline\noalign{\smallskip}
\multicolumn{3}{c}{December 2009}\\
\hline
13 & XL 2.16-m & 5.18 \\
14 & XL 2.16-m & 4.86 \\
15 & XL 2.16-m & 7.15 \\
16 & XL 2.16-m & 7.09 \\
17 & XL 2.16-m & 7.10 \\
18 & XL 2.16-m & 5.49 \\
25 & LJ 2.40-m & 4.61 \\
26 & LJ 2.40-m & 5.90 \\
27 & LJ 2.40-m & 6.00 \\
28 & LJ 2.40-m & 8.18 \\
29 & LJ 2.40-m & 2.00 \\
30 & LJ 2.40-m & 4.36 \\
31 & LJ 2.40-m & 2.52 \\
\hline
\multicolumn{3}{c}{January 2010}\\
\hline
12 & SPM 1.5-m & 5.09 \\
~ & XL 2.16-m & 4.14 \\
13 & SPM 1.5-m & 5.43 \\
~ & XL 2.16-m & 3.04 \\
14 & SPM 1.5-m & 5.10 \\
~ & XL 2.16-m & 4.42 \\
15 & SPM 1.5-m & 5.11 \\
~ & XL 2.16-m & 4.58 \\
16 & SPM 1.5-m & 3.93  \\
~ & XL 2.16-m & 4.29  \\
17 & SPM 1.5-m & 5.13  \\
~ & XL 2.16-m & 4.29  \\
24 & LJ 2.40-m & 5.63 \\
26 & LJ 2.40-m & 6.10 \\
27 & LJ 2.40-m & 1.75 \\
28 & LJ 2.40-m & 2.97 \\
29 & LJ 2.40-m & 5.38 \\
30 & LJ 2.40-m & 6.05 \\
31 & LJ 2.40-m & 5.90 \\
\noalign{\smallskip}\hline
  \end{tabular}\end{center}
\end{table}

The CCD data were reduced with the iraf Daophot package. The PMT data were 
reduced as described in Pfeiffer et al. (1996). Unfortunately the weather
conditions were poor at Piszk\'estet\H{o} Observatory during the campaign
in 2007, so these data were not included in the final light curves of the year.

The data collected in both BOAO and XL in December 8 and 11 overlapped
partially each other. As they were obtained through different filters,
Johnson $B$ for BOAO and white light for XL, the overlapping data were
used to calculate the amplitude ratio obtained between the two time-series.
Then, the
magnitudes of the light curves from XL were calibrated to those in the filter
$B$. Hence, data from the two sites were able to be combined together for
frequency analysis. Since the noise level of the PMT data was in general
higher than that of the CCD data, the overlapping PMT data were not used for
the analysis of the light curves. Fig.~1(a) and (b) show the calibrated light
curves of HS~0507+0434B in 2007 and 2009-2010, respectively.

\begin{figure}
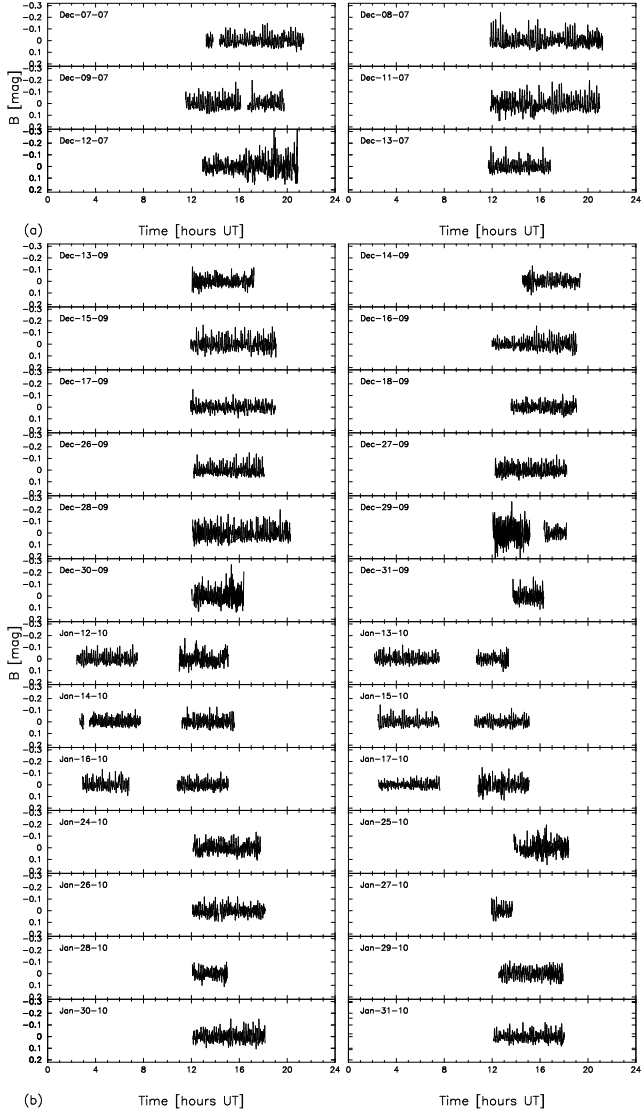

\resizebox{\hsize}{!}{\includegraphics{fig1a.ps}}
\vspace{3mm}
\resizebox{\hsize}{!}{\includegraphics{fig1b.ps}}
\caption{Normalized light curves for HS~0507+0434B in (a) December 7-13, 2007; (b) from December 13, 2009 to January 31, 2010. Each panel covers a 24~h period. 
The date is indicated on the up-left side inside the panels.}
\label{fig1}
\end{figure}

\section{The amplitude spectrum}

We used the {\tt PERIOD04} software (Lenz \& Breger 2005) to analyse the
light curves and derive the frequencies, amplitudes and phases of the peaks
in the Fourier Transforms. The amplitude spectra of the light curves in 2007
and 2009-2010 are shown in the upper panels of Fig.~2(a) and (b), respectively. 
The amplitude spectra of a sinusoidal function sampled at the same rate as the
data and having an amplitude equals to unity when there are data and equals to
zero in the gaps (the window functions) are shown in the insets of upper panels
of Fig.~2(a) and (b).

\begin{figure}
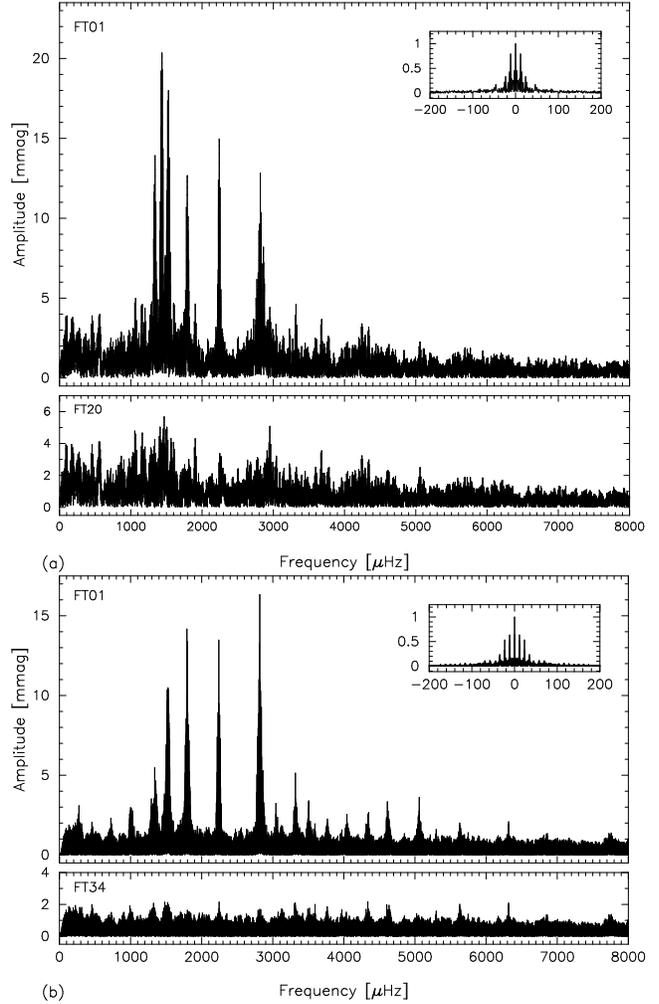

\resizebox{\hsize}{!}{\includegraphics{fig2a.ps}}
\vspace{10mm}
\resizebox{\hsize}{!}{\includegraphics{fig2b.ps}}
\caption{Amplitude spectra of the light curves and the residuals after
prewhitenning the detected frequencies in (a) 2007; and (b) 2009-2010
shown in the upper and bottom panels, respectively. The amplitude, in units
of milli-magnitude amplitude (mmag), is shown as a function of the
frequency in $\mu$Hz in the frequency range 0-8000~$\mu$Hz; outside this
frequency range no significant signal is detected. The insets are the spectral
windows of the data. Please note that the insets and the figures are in
different scales.}
\label{fig2}
\end{figure}

From the Fourier Transform (FT) of the light curves in 2007, we extracted
frequency values of the highest peaks with {\tt PERIOD04}, fitting the
light curves with the frequencies then prewhitenning the corresponding
sinusoidal function to look for the next frequencies with significant
peaks in the FT. For the peaks extracted, we followed the criterion of
Breger et al. (1993) and Kuschnig et al. (1997) to take the peaks whose
signal-to-noise (S/N) ratios are higher than 4.0. Finally, 19 significant
frequencies were detected from the light curves collected in 2007.
The first two columns of Table~3 list the resolved frequencies and their
amplitudes in 2007. In the bottom panel of Fig.~2(a) is shown the FT of
the residuals after prewhitenning the 19 frequencies.

For the light curves obtained from December of 2009 to January of 2010, we note
that the amplitude spectra of the four weeks of light curves differ 
significantly from each other, especially for the amplitude values of the peaks.
Fig.~3 shows the FTs of the light curves of the four individual weeks
(namely Week1, Week2, Week3, and Week4, respectively). We discuss those amplitude 
variations in more details in $\S 5$.

To extract the frequency values from the FT of the combined light curves from
December of 2009 to January of 2010, we made an analysis similar to the one
used  to analyse the light curves of 2007. However, since the amplitude of a
given peak is varying from week to week, prewhitenning of each of the peak by
one frequency with a unique fitted amplitude and phase does not allow to
completely remove that sinusoidal function for all the four weeks of light
curves. The existence of a residual of the prewhitenning by a sinusoidal
function creates a secondary peak whose frequency value is very close to that
of the resolved frequency with a lower amplitude. In order to make complete
prewhitenning, we optimized the amplitude of each frequency for individual
weeks of light curves and then prewhitenned the corresponding sinusoidal
functions with varying amplitude values. This process helps us to extract 33
frequencies, that are listed in the third column of Table~3 while the fourth
column gives the average amplitude values. The FT of the residuals after
prewhitenning by the 33 frequencies is shown in the bottom panel of Fig.~2(b).
We used the same selection  criterion of S/N ratios higher than 4.0 as for the
analysis of the 2007 data.
 As the uncertainties derived by {\tt PERIOD04}
estimate the internal consistency of the solutions, they are believed to be
underestimated. We derived more realistic uncertainties through Monte-Carlo
simulations.

For each observed time-series $(t_i$, $x_i)_{i=1,2,...,k}$, we extracted a
number of frequencies $(f_j)_{j=1,2,...,n}$ and the corresponding amplitudes
$A_j$ and phases $\phi _j$ with the software {\tt PERIOD04}, so the residual
time-series $(t_i, y_i)_{i=1,2,...,k}$ were obtained by subtracting the sum
of multiple sine functions from the observed time-series as

$$
y_i=x_i-\sum_{j=1,2,...n} A_j\sin (2\pi f_jt_i+\phi _j)
$$

Then $|y_i|$ is regarded as the observation error of $x_i$. Hence, we
constructed 50 simulated time-series $z_i$ for each observed time-series, by
taking $z_i$ as a normally distributed random variable with mean $x_i$ and
standard variation $|{y_i}|$. The 50 simulated time-series were fitted with
the sum of multiple sine functions by taking $(f_j, A_j, \phi _j)_{j=1,2,...,n}$
as initial values according to least-squares algorithm. Hence 50 sets of new
$(f_j, A_j, \phi _j)_{j=1,2,...,n}$
were obtained. The standard deviations of each parameter of
$(f_j,A_j,\phi _j)_{j=1,2,...,n}$ were then calculated. We define them as the
uncertainty estimate of the parameters and include those of frequencies and
amplitudes in Table~3.

 We compared the uncertainties derived from Monte-Carlo simulations with those derived from  {\tt PERIOD04}
in order to estimate by how much {\tt PERIOD04} underestimates those uncertainties. For the 2007 data set, the 
 average uncertainties on the frequencies and on the amplitudes derived from our Monte-Carlo
simulations are  70\% larger than those estimated from {\tt PERIOD04}. For the 2009-2010 data sets, 
{\tt PERIOD04} underestimates the frequency uncertainties by 81\% and the amplitude uncertainties by 48\%, in average.

\begin{figure}
\resizebox{\hsize}{!}{\includegraphics{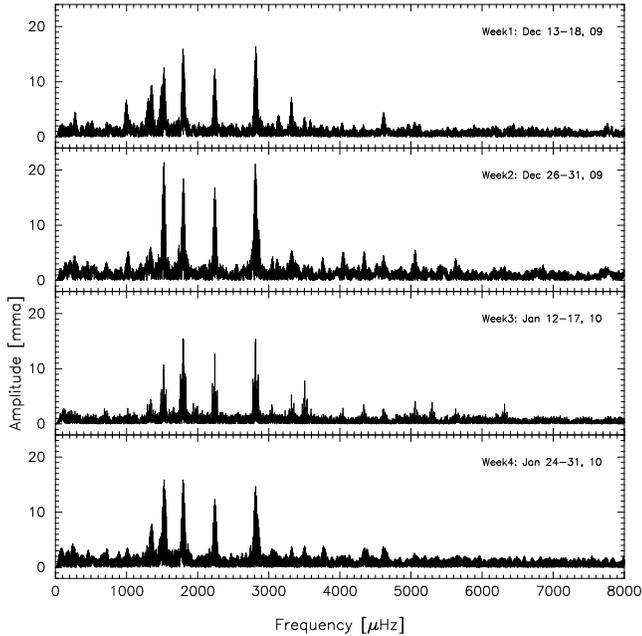}}
\caption{Amplitude spectra of the four weeks of light curves from December
 2009 to January  2010. Note the significant differences of the
amplitudes of the peaks between the four weeks.}
\label{fig3}
\end{figure}

\begin{table}
\footnotesize
\tabcolsep=0.8mm
\begin{center} \caption{Signals detected from the light curves of HS~0507+0434B
in 2007 and 2009-2010. $f$=frequency in $\mu$Hz. $A$=amplitude in mmag.
The 1$\sigma$ uncertainties are estimated from Monte-Carlo simulations. }
\begin{tabular}{rrrr}
\hline
\multicolumn{1}{c}{$f$} &\multicolumn{1}{c}{$A$} &\multicolumn{1}{c}{$f$} & \multicolumn{1}{c}{$A$}\\
\multicolumn{2}{c}{(2007)} & \multicolumn{2}{c}{(2009-2010)}\\
\hline
&&   274.983 $\pm$ 0.018 & 3.12 $\pm$ 0.44\\
&&   723.231 $\pm$ 0.024 & 2.34 $\pm$ 0.55\\
&&  1000.262 $\pm$ 0.019 & 3.04 $\pm$ 0.57\\
&&  1028.558 $\pm$ 0.035 & 2.83 $\pm$ 0.45\\
&&  1292.371 $\pm$ 0.015 & 3.90 $\pm$ 0.62\\
1332.797 $\pm$ 0.099 & 10.86 $\pm$ 1.14 &&\\
1335.831 $\pm$ 0.148 & 10.51 $\pm$ 1.17 &&\\
1340.241 $\pm$ 0.078 & 12.98 $\pm$ 0.82 & 1340.716 $\pm$ 0.011 & 4.50 $\pm$ 0.64\\
&&  1351.176 $\pm$ 0.046 & 2.55 $\pm$ 0.41\\
&&  1355.893 $\pm$ 0.008 & 4.16 $\pm$ 0.54\\
1420.664 $\pm$ 0.240 & 9.06 $\pm$ 1.15 &&\\
1429.478 $\pm$ 0.075 & 15.19 $\pm$ 1.05 &&\\
1437.551 $\pm$ 0.053 & 17.43 $\pm$ 1.04 &&\\
&&  1476.397 $\pm$ 0.012 & 3.38 $\pm$ 0.54\\
&&  1519.086 $\pm$ 0.019 & 5.32 $\pm$ 0.50\\
&&  1521.827 $\pm$ 0.004 & 10.58 $\pm$ 0.56\\
1524.484 $\pm$ 0.074 & 12.67 $\pm$ 1.02 & 1524.559 $\pm$ 0.015 & 5.84 $\pm$ 0.59\\
1528.924 $\pm$ 0.042 & 24.13 $\pm$ 1.50 & 1527.296 $\pm$ 0.006 & 11.04 $\pm$ 0.64\\
&&  1529.134 $\pm$ 0.012 & 4.66 $\pm$ 0.53\\
1541.309 $\pm$ 0.090 & 13.61 $\pm$ 1.39 &&\\
&& 1549.485 $\pm$ 0.072 & 2.76 $\pm$ 0.44\\
1792.277 $\pm$ 0.073 & 14.93 $\pm$ 0.87 & 1792.809 $\pm$ 0.003 & 14.95 $\pm$ 0.58\\ 
1796.882 $\pm$ 0.219 & 5.03 $\pm$ 0.99 & 1796.841 $\pm$ 0.010 & 6.73 $\pm$ 0.51\\ 
1801.518 $\pm$ 0.105 & 10.87 $\pm$ 0.92 & 1800.841 $\pm$ 0.005 & 14.26 $\pm$ 0.63\\
&& 1802.138 $\pm$ 0.057 & 2.07 $\pm$ 0.39\\
2241.497 $\pm$ 0.069 & 14.71 $\pm$ 1.10 & 2241.642 $\pm$ 0.004 & 13.83 $\pm$ 0.45\\
&& 2248.724 $\pm$ 0.014 & 5.29 $\pm$ 0.51\\
2770.036 $\pm$ 0.142 & 6.85 $\pm$ 1.10 &&\\
2810.513 $\pm$ 0.112 & 8.46 $\pm$ 0.98 & 2810.768 $\pm$ 0.006 & 12.55 $\pm$ 0.46\\
&& 2814.219 $\pm$ 0.028 & 2.51 $\pm$ 0.52\\
2817.649 $\pm$ 0.076 & 12.30 $\pm$ 0.88 & 2817.728 $\pm$ 0.004 & 15.81 $\pm$ 0.59\\
2859.908 $\pm$ 0.164 & 6.39 $\pm$ 1.00 &&\\
2861.801 $\pm$ 0.153 & 8.26 $\pm$ 0.91 & 2861.726 $\pm$ 0.017 & 4.21 $\pm$ 0.50\\
&& 3043.694 $\pm$ 0.018 & 3.34 $\pm$ 0.47\\
3318.634 $\pm$ 0.178 & 4.64 $\pm$ 0.97 & 3318.685 $\pm$ 0.014 & 5.10 $\pm$ 0.59\\
&& 3505.943 $\pm$ 0.037 & 3.47 $\pm$ 0.54\\
&& 3768.946 $\pm$ 0.024 & 2.32 $\pm$ 0.39\\
&& 4042.533 $\pm$ 0.041 & 2.53 $\pm$ 0.41\\
&& 4349.907 $\pm$ 0.026 & 2.66 $\pm$ 0.50\\
&& 4610.526 $\pm$ 0.018 & 3.28 $\pm$ 0.51\\
&& 5059.364 $\pm$ 0.023 & 3.60 $\pm$ 0.51\\
\hline
\end{tabular}
\end{center}
\end{table}

\section{Asteroseismology of HS0507+0434B}

\subsection{Pulsation frequencies and linear combinations}
As shown in Table 1, the useful data for HS 0507+0434B obtained in 2007 were collected
within one week from XL and BOAO, which are only one hour apart
in longitude. This leads to a strong one-day aliasing effect 
in the FT, as seen in the window function shown in the inset of Fig.~2(a).
For the four-week data-set collected from December 2009 to January 2010 (Table 2),
there are three weeks of observations made from single sites. Hence, the
one-day aliasing peaks persist in the FT, as seen in the window function shown in the inset of Fig.~2(b),
but with a smaller 11.5~$\mu$Hz component.

Since  HS~0507+0434B lies in the middle of the ZZ Ceti instability strip, 
where the convection-pulsation interaction starts to become important, 
we anticipated to find  many linear combinations of a few independent 
frequencies as in Handler et al. (2002). We carefully searched for independent 
frequencies and their linear combinations, taking into account possible remaining  aliasing effect
and residuals of amplitude variation on the frequency determination.

Table~4 lists the frequencies, amplitudes and periods resulting from our
analysis. When the same frequencies occur in the both sets of data (2007 and
2009-2010), we choose the 2009-2010 values since the frequency resolution is
better for this data set. In this case, the listed amplitude is the amplitude
averaged on the four weeks of the 2009-2010 runs. We searched systematically
for linear combination relationship among three frequencies A,B and C such that:
 
$f_{A}$$\pm$$\sigma_{A}$= $f_{B}$$\pm$$\sigma_{B}$ $\pm$ $f_{C}$$\pm$$\sigma_{C}$

We select as possible linear combinations only those combinations with: 

$\sigma_{A}$ $\leq$ 3$\times$[$\sigma_{B}$ + $\sigma_{C}$]

 For every linear
combination found, we identify the higher amplitude peaks as independent
signals and the lowest one as the combination. We also list as further signals, a number of
frequencies of low amplitude which are more uncertain detections at the
limit of our selection criterion and/or which are close to possible linear combinations but do not fulfill our 
3$\times$$\sigma$ criterion. 
Among those frequencies, we discuss two particular cases as follows.
 The 723.23~$\mu$Hz which we identify here as a linear combination $f_{13}$-$f_{23}$ within 3$\times$$\sigma$ 
could also correspond to a genuine mode. This will be discussed in section 6.
The 1292.37~$\mu$Hz that is listed as a linear combination does not fulfill our amplitude selection criterion since its presumably parent
mode $f_{17}$ has a smaller amplitude. However it fulfills the frequency criterion within 0.6$\times$$\sigma$. We suggest that the linear combinations
 between the largest amplitude components of the implied two triplets( $f_{16}$ and  $f_{18}$ interacting with $f_{7}$ and  $f_{9}$) combine to produce this almost exact 
 linear combination.

 In Table~4, we do not include 1541.31~$\mu$Hz
which was only seen in the 2007 data since this peak is regarded as the
one-day alias of the dominant peak at 1528.92~$\mu$Hz.
  Similarly, we exclude the 1351.17~$\mu$Hz peak as the alias of the 1340.71~$\mu$Hz. 
We also exclude the peak at 
1802.14~$\mu$Hz since it may result from the residual of the
amplitude variation of the peak at 1800.84~$\mu$Hz. We include the 2245.63~$\mu$Hz from Handler et al. (2002)
for completeness. For these reasons, Table 4 contains 39 frequencies while Table 3 contains 41.

In comparing with the results of Handler et al. (2002), we find in common the same two complete triplets centered on
 1796~$\mu$Hz ($f_{10}$,$f_{11}$ and $f_{12}$) and 2814~$\mu$Hz ($f_{16}$, $f_{17}$ and $f_{18}$), the two $m$= $\pm$1
 components of the triplet centered at 2245~$\mu$Hz ($f_{13}$ and $f_{15}$; we do not detect the central component of that triplet
in our data but we include the Handler's value in Table 4 as $f_{14}$ for completeness)
 and the triplet centered on 1335~$\mu$Hz ( $f_2$,  $f_3$ and $f_4$) where Handler et al. (2002) detected only one component. In addition we find one more
triplet centered on 1524~$\mu$Hz ( $f_7$,  $f_8$ and $f_9$) and two frequencies ($f_5$ and $f_6$) separated by twice the averaged separation between the components 
of complete triplets.
We interpret these two frequencies as the $m$= $\pm$1 components of an additional triplet whose undetected central component frequency 
 should be at 1433.51~$\mu$Hz. The lowest frequency $f_1$ must be a genuine mode since we do not identify any linear combination 
reproducing this value. By contrast with Handler's results, we do not find as many linear combination
frequencies satisfying our S/N selection criterion.

\begin{table}
\footnotesize
\begin{center} \caption{Signals identified for HS~0507+0434B. $f$=frequency in $\mu$Hz. $A$=amplitude in mmag. P=Period in second.}
\begin{tabular}{crrr}
\hline
ID & \multicolumn{1}{c}{$f$} & \multicolumn{1}{c}{$A$} & \multicolumn{1}{c}{P} \\
\hline
\multicolumn{4}{c}{Independent signals}\\
\hline
   $f_1$ &  1000.26 &  3.04 & 999.7 \\
   $f_2$ &  1332.80 & 10.86 & 750.3 \\
   $f_3$ &  1335.83 & 10.51 & 748.6 \\
   $f_4$ &  1340.24 & 12.98 & 746.1 \\
   $f_5$ &  1429.48 & 15.19 & 699.6 \\
   $f_6$ &  1437.55 & 17.43 & 695.6 \\
   $f_7$ &  1521.83 & 10.58 & 657.1 \\
   $f_8$ &  1524.56 &  5.84 & 655.9 \\
   $f_9$ &  1527.30 & 11.04 & 654.8 \\
$f_{10}$ &  1792.81 & 14.95 & 557.8 \\
$f_{11}$ &  1796.84 &  6.73 & 556.5\\
$f_{12}$ & 1800.84 & 14.26 & 555.3\\ 
$f_{13}$ & 2241.64 & 13.83 & 446.1\\  
$f_{14}$ & $^a$2245.63 & 2.8 & 445.3\\  
$f_{15}$ & 2248.72 & 5.29 & 444.7\\    
$f_{16}$ & 2810.77 & 12.55 & 355.8\\  
$f_{17}$ & 2814.22 & 2.51 & 355.3\\  
$f_{18}$ & 2817.73 & 15.81 & 354.9\\ 
\hline
\multicolumn{4}{c}{Further signals}\\
\hline
$f_{19}$ & 1028.56 & 2.83 & 972.2 \\
$f_{20}$ & 1355.89 & 4.16 & 737.5 \\
$f_{21}$ & 1420.66 &  9.06 & 703.9\\
$f_{22}$ & 1476.40& 3.38&677.3\\
$f_{23}$ & 1519.09 & 5.32 & 658.3 \\
$f_{24}$ & 1529.13 & 4.66 & 654.0 \\
$f_{25}$ & 1549.48 & 2.76 & 645.4 \\
$f_{26}$ & 2861.80&8.26&349.4\\  
$f_{27}$ & 3505.94 & 3.47 & 285.2 \\
$f_{28}$ & 4349.91 & 2.66 & 229.9 \\
\hline
\multicolumn{4}{c}{Linear combinations}\\
\hline
$f_{11}-f_{7}$& 274.98& 3.12&3636.6\\
$f_{13}-f_{23}$& 723.23& 2.34&1382.7\\
$f_{17}-f_{7}$&1292.37& 3.90&773.8\\
$f_{2}+f_{6}$&2770.04&6.85&361.0\\   
$f_{3}+f_{8}$&2859.91&6.39&349.7\\  
$f_{7}+f_{7}$&3043.69&4.34&328.5\\
$f_{7}+f_{11}$&3318.68 & 5.10 & 301.3\\ 
$f_{9}+f_{13}$&3768.95&2.32&265.3\\  
$f_{12}+f_{13}$&4042.53 & 2.53 & 247.4\\ 
$f_{10}+f_{18}$&4610.53&3.28&216.9\\
$f_{13}+f_{18}$&5059.36 & 3.60 & 197.7\\ 
\hline
\multicolumn{4}{l}{$^a$ From Handler et al. (2002).}\\
\end{tabular}
\end{center}
\end{table}

\subsection{Period distribution and average period spacing}
From Table~4, one finds that the 18 independent signals consist of 6
triplets (with the one central mode missing in the triplet between $f_5$ and $f_6$ as mentioned earlier)
and 1 single mode. Since we do identify only triplets in the power spectrum, we interpret them as $l=1$ modes split by rotation.
In the asymptotic regime, which we assume here in first approximation, the period separation between modes of same degree $l$ and consecutive orders $k$
should be equal or close to the period spacing. Hence, the observed period
separations should correspond to the period spacing, or be approximately
its multiples. Examining the periods of the central modes in the triplets and of
the single mode, one finds the differences of 251.1~s, 51.0~s, 41.7~s, 99.4~s,
111.2~s, and 90.0~s, respectively. Since the $m$
value of the single mode is not determined, we used the periods of
the central modes of the triplets to calculate the average period spacing.
This leads to the average period spacing of $\Delta P$=49.63~s.

The period distribution of the $m=0$ modes allows us to assign an arbitrary $\delta k$ value to the
$m=0$ modes taking the frequency 2814.22~$\mu$Hz (period 355.3~s) as the reference $\delta k$=0.
 Table~5 lists the frequencies, the frequency separations, the periods,
the $\delta k$ and the $m$ values for the identified  $l$=1 modes.
A straight-line fit to the periods of the six $m=0$ modes yields the formula,

$$
{\rm P}_k=P_0 + \Delta P \times \Delta k
$$

where one gets $P_0$=353.18$\pm$4.12~s and $\Delta P$=49.63$\pm$0.78~s. Hence,
${\rm DP}$=P-P$_k$ are derived for the six $m=0$ modes and listed in Table~5.

Fig.~4 shows the linear least-square fit of the six $m=0$ modes, which confirms 
the period spacing of $\Delta P$=49.63~s. 
 In addition, by checking the periods listed in Table~4, one finds
that the differences between the period of $f_{17}$ (whose $m$ value is 0) and the periods corresponding to the three
frequencies of 3318.68, 4042.53, and 5059.36~$\mu$Hz (301.3~s, 247.4~s and 197.7~s respectively), which we interpret as linear combinations,
are 54.0, 53.9, and 49.7~s, respectively. Hence, the periods of these three linear combinations 
 fit surprisingly well the period spacing of 49.63~s. One may wonder whether these linear combination frequencies coincide with genuine $l=1$ modes of lower order.
We will discuss this point further in $\S 6$.

\begin{table}
\footnotesize
\tabcolsep=1.2mm
\begin{center} \caption{Period distribution in HS~0507+0434B. $f$=frequency in $\mu$Hz. P=Period in second. $\delta f$=frequency separation among a triplet in $\mu$Hz.
 DP are the residuals of the fit to the periods of the six $m=0$ modes. 
 The frequency indicated in $italic$ is estimated from the average frequency of the two components $m$=$\pm$1 of the triplet.}
\begin{tabular}{rrrrrr}
\hline
\multicolumn{1}{c}{$f$} & \multicolumn{1}{c}{$\delta f$} & \multicolumn{1}{c}{P} & \multicolumn{1}{c}{$\delta k$} & \multicolumn{1}{c}{$m$} & \multicolumn{1}{c|}{DP}\\
\hline
1332.80 &  & 750.3 &  & -1 &\\
       &3.03&&&&\\
1335.83 &  & 748.6 & +8 &  0 &-1.65\\
       &4.41&&&&\\
1340.24 &  & 746.1 &  & +1 &\\
1429.48 &  & 699.6 &  & -1 &\\
       &4.03&&&&\\
\it{1433.51} & & \it{697.6} & \it{+7} & \it{0} &-3.02\\
       &4.03&&&&\\
1437.55 &  & 695.6 &  & +1 &\\
1521.83 &  & 657.1 &  & -1 &\\
       &2.73&&&&\\
1524.56 &  & 655.9 & +6 &  0 &4.92\\
       &2.74&&&&\\
1527.30 &  & 654.8 &  & +1 &\\
 1792.81 &  & 557.8 &  & -1 &\\
 &4.03&&&&\\
 1796.84 &  & 556.5 & +4 &  0 &4.78\\
 &4.00&&&&\\
 1800.84 &  & 555.3 &  & +1 &\\
 2241.64 &  & 446.1 &  & -1 &\\
 &3.99&&&&\\
 2245.63 &  & 445.3 & +2 &  0 &-7.15\\
 &3.09&&&&\\
 2248.72 &  & 444.7 &  & +1 &\\
 2810.77 &  & 355.8 &  & -1 &\\
 &3.45&&&&\\
 2814.22 &  & 355.3 & 0 &  0 &2.12\\
 &3.51&&&&\\
 2817.73 &  & 354.9 &  & +1 &\\
\hline
\end{tabular}
\end{center}
\end{table}

\subsection{Rotational splitting and rotation rate}
We interpret the 6 triplets identified in the FT as $l=1$ modes split by rotation. Column 2 of Table~5 lists the
frequency separation  between the rotationally split components, $m$= $\pm$1,  and the central component, $m$=0,  for the 6 triplets. 
We derive an average value of frequency separation between the outer two
components of the triplets as {\bf 3.59$\pm$0.57~$\mu$Hz.} To the first order in the rotation angular frequency $\Omega$,
the frequencies in the rotating case $\sigma_{k,l,m}$ are related to the
frequencies in the non-rotating case $\sigma_{k,l}$ by:

$$
\sigma_{k,l,m}=\sigma_{k,l}+m\times (1-C_{k,l}) \Omega
$$

with $C_{k,l}=1/l(l+1)$ in the asymptotic regime (Brickhill 1975). Assuming
that it is the case of the pulsations observed in HS~0507+0434B, we derive
an average rotation period of {\bf 1.61$\pm 0.26$~days.} This value is in
agreement with the rotation period derived by Handler et al. (2002) of
1.70$\pm 0.11$~days within the uncertainties.

\begin{figure}
\resizebox{\hsize}{!}{\includegraphics{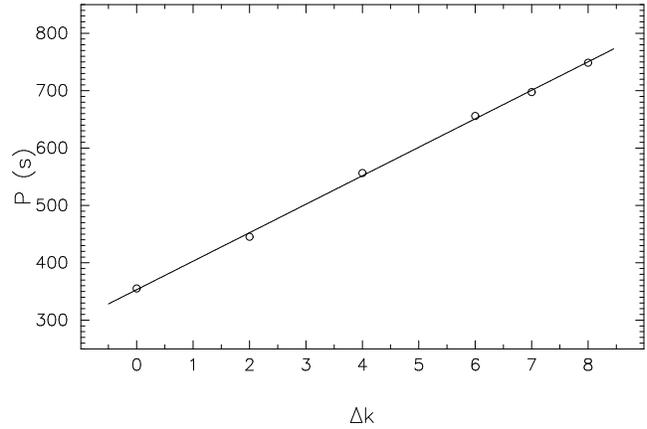}}
\caption{Linear least-square fit of the six $m=0$ modes (open circles) as a function of $\Delta k$, the differences in radial order from the reference mode. 
The fit confirms a period spacing $\Delta P$=49.63~s. }
\label{fig4}
\end{figure}

\subsection{The angle of the rotation axis to the line of sight}
Among the five triplets (the sixth one with central mode missing is not included)
listed in Table~4, one may find that four of them show weaker central modes
($m=0$) obviously than the $m=-1$ and $m=+1$ modes in the same triplets.
Making the assumptions that the pulsation axis is aligned with the star rotation axis and that
the average amplitudes ratios of the components within the multiplets are due to geometrical effects only, it is 
possible to infer the angle of the rotation axis on the line of sight.
We hence calculate the ratio of the average amplitudes of the $m=\pm 1$
to that of the amplitudes of the $m=0$ modes for the four triplets. The
result is 1.98. This implies that the angle of the rotation axis to the line
of sight is close to $70\,^{\circ}$ (Pesnell 1985). However, we note that 
the fifth triplet centered on 
1335.8~$\mu$Hz has its three components of similar amplitude. Either these particular pulsation modes have
a different symmetry axis than the other four triplets, or the amplitude
ratios are not only due to geometrical effect.  
The full triplet was seen in 2007 while only the prograde component was detected in 2009-2010 with an amplitude reduced by a factor 3.
Those amplitude variations, seen for other modes as well, are discussed below in $\S 5$. They suggest that the amplitude ratio cannot be  simply explained by 
geometrical effect.

\subsection{Mode trapping}
With the linear least-square fit of the periods of the six $m=0$ modes, we
calculate the residuals of these six periods relative to the linear fit with the average period spacing of 49.63~s.
Fig.~5 plots the residuals versus the periods, which shows evidence of
mode trapping.
 The two modes with periods of 445.3~s and 697.6~s show the largest deviations from the average period spacing.
 Both shoud be trapped modes or close to trapped modes. They are separated in periods by five times the average period spacing, i.e. 
five modes. This is a preliminary indication of a trapping  by a thin hydrogen layer. Brassard et al. (1992)
 and Kawaler \& Bradley (1994) have shown that the distance in periods, and consequently the number of modes, 
 between trapped modes increases as the surface hydrogen layer becomes thinner.
In the mean time, from the linear theory of pulsation, the trapped modes are expected to have the largest amplitudes since they have the largest growth-rates.
 However, this has been shown not to be true in many cases: for instance in PG~1159-035 (Winget et al. 1991), in GD~358 (Winget et al. 1994),
 in RXJ~2117+3412 (Vauclair et al. 2002), in EC~14012-1446 (Provencal et al. 2012). In HS~0507+0434B as in those other pulsators, 
there is no correlation between the amplitudes of the modes  
and their trapped or untrapped status, indicating that the amplitudes are not simply governed by their linear growth-rate. Amplitude and frequency variations, as
discussed below in section 5, may be the reason why the triplet at 1433~$\mu$Hz is not detected in the 2009-2010 data set in spite of being a probable trapped mode.

\begin{figure}
\resizebox{\hsize}{!}{\includegraphics{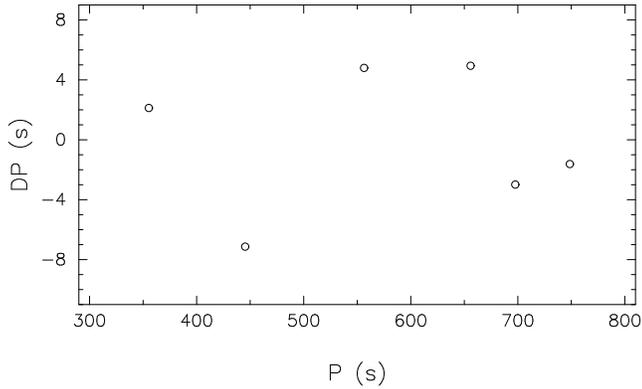}}
\caption{A plot showing observational evidence of mode trapping. The residuals of the period distribution relative to the average period spacing ($\Delta P$=49.63~s) 
are plotted as a function of the period. The open circles represent the $m=0$ modes of the six triplets. The modes
with periods of 445.5~s and 697.6~s defining  clear minima on this plot should be trapped modes.}
\label{fig5}
\end{figure}

\section{The amplitude variation}
By comparing the upper panel of Fig.~2(a), the four panels of Fig.~3, and
Fig.~3 of Handler et al. (2002) based on one-week single-site CCD photometric
observations in 2000, one finds quite different amplitude spectra. As
mentioned in $\S 3$, we had to optimize the amplitude values of the same
frequencies for the four different weeks of light curves from December 
2009 to January  2010 when we extracted frequencies from the FT of the
combined light curves.
Fig.~6 shows the amplitude variations of the 33
frequencies. Table~6 lists the 33 frequencies and their
amplitudes during  the four individual weeks. 
 We checked that for each of the four weeks,
 the achieved frequency resolution allows to resolve the rotational splitting. The 
frequency resolution is 2.2~$\mu$Hz, 1.9~$\mu$Hz, 2.1~$\mu$Hz and 1.6~$\mu$Hz for week1,week2, week3 and week4 respectively
while the rotational splitting is 3.6~$\mu$Hz. This indicates that the observed amplitude variations are real and 
are not due to the beating between the components
within the triplets. 

The ``pulsation power'' of HS~0507+0434B
during the four individual weeks is estimated in
$mmag^2$ per second by computing:

$$
E=\displaystyle\sum\limits_{i=1}^{33} f_i \times A_i^2
$$

The results are listed in the bottom of Table~6.
A similar estimate for the 2007 data set gives $E$= 5.11~$mmag^2$ per second.
These results suggest that the ``pulsation power'' is time-dependent. HS~0507+0434B was 
pulsating more vigorously during our 2007 run and during the week2 of December 2009 than during the other
three weeks of December 2009 and January 2010 where the  ``pulsation power'' was almost constant.

Amplitude variations are not uncommon among pulsating white dwarfs. They have been noticed, for instance, in the ZZ Ceti pulsators KUV~02464+3239
(Bogn\'{a}r et al. 2009), EC~14012-1446 (Provencal et al. 2012), WDJ~1916+3938 (Hermes et al. 2011), in the DBV prototype GD~358 (Kepler et al. 2003, Provencal et al. 2009)
as well as in the GW Virginis star PG~0122+200 (Vauclair et al. 2011).
 Such amplitude variations associated with frequency variations are expected in the case of resonant coupling induced by the rotation within multiplets 
(Goupil et al. 1998). The interaction of the pulsation with convection could also be invoked as the cause for the amplitude and  ``pulsation power'' variability.
 More observational data and theoretical investigations  are needed to explore the physical causes of those variabilities. This is out of the scope of the present paper.

\begin{table}
\footnotesize
\tabcolsep=1.2mm
\begin{center} \caption{Frequencies extracted from the combined light curves from December 2009 to January  2010 and the amplitudes optimized for the 
four individual weeks. $f$=frequency in $\mu$Hz. $A$=amplitude in mmag. $E$ represents the pulsation power in $mmag^2$ per second of the star in the four 
individual weeks (see the text for the details).}
\begin{tabular}{rrrrrr}
\hline
&&Week1&Week2&Week3&Week4\\
\cline{3-6}
\multicolumn{1}{c}{ID} & \multicolumn{1}{c}{$f$} & \multicolumn{4}{c}{$A$}\\
\hline
$F_{01}$ &   274.98 &    3.27 &    3.57 &    2.23 &    3.38\\
$F_{02}$ &   723.23 &    2.66 &    1.06 &    2.44 &    3.46\\
$F_{03}$ &  1000.26 &    5.42 &    4.05 &    1.03 &    3.11\\
$F_{04}$ &  1028.56 &    3.05 &    4.70 &    1.88 &    1.16\\
$F_{05}$ &  1292.37 &    5.42 &    4.02 &    3.53 &    3.43\\
$F_{06}$ &  1340.72 &    5.80 &    3.43 &    3.23 &    5.04\\
$F_{07}$ &  1351.18 &    3.60 &    2.47 &    2.34 &    5.83\\
$F_{08}$ &  1355.89 &    9.20 &    3.62 &    0.49 &    7.28\\
$F_{09}$ &  1476.40 &    7.12 &    2.09 &    1.94 &    4.26\\
$F_{10}$ &  1519.09 &    2.20 &    5.93 &    5.12 &    6.91\\
$F_{11}$ &  1521.83 &    8.50 &    9.64 &   10.32 &   14.19\\
$F_{12}$ &  1524.56 &    2.29 &    5.39 &    4.64 &    7.43\\
$F_{13}$ &  1527.30 &    9.82 &   20.12 &    1.23 &   11.97\\
$F_{14}$ &  1529.13 &    6.94 &    1.87 &    6.05 &    6.17\\
$F_{15}$ &  1549.48 &    2.41 &    3.23 &    3.36 &    7.86\\
$F_{16}$ &  1792.81 &   16.93 &   13.97 &   14.22 &   16.81\\
$F_{17}$ &  1796.84 &    6.70 &    8.18 &    5.27 &    5.08\\
$F_{18}$ &  1800.84 &   12.05 &   19.95 &   15.54 &   10.34\\
$F_{19}$ &  1802.14 &    3.91 &    2.21 &    2.95 &    4.54\\
$F_{20}$ &  2241.64 &   12.84 &   16.41 &   13.12 &   12.43\\
$F_{21}$ &  2248.72 &    4.80 &    8.63 &    5.49 &    3.07\\
$F_{22}$ &  2810.77 &    9.07 &   20.16 &   15.04 &    6.35\\
$F_{23}$ &  2814.22 &    2.08 &    3.10 &    2.37 &    3.36\\
$F_{24}$ &  2817.73 &   15.49 &   18.23 &   15.77 &   14.28\\
$F_{25}$ &  2861.73 &    4.07 &    5.82 &    2.37 &    4.69\\
$F_{26}$ &  3043.69 &    2.81 &    4.03 &    3.30 &    3.26\\
$F_{27}$ &  3318.68 &    6.75 &    5.50 &    5.36 &    3.58\\
$F_{28}$ &  3505.94 &    3.11 &    1.97 &    7.54 &    2.44\\
$F_{29}$ &  3768.95 &    1.57 &    3.24 &    1.66 &    2.81\\
$F_{30}$ &  4042.53 &    1.32 &    4.80 &    2.86 &    0.91\\
$F_{31}$ &  4349.91 &    0.60 &    4.96 &    1.72 &    3.06\\
$F_{32}$ &  4610.53 &    4.36 &    4.04 &    2.29 &    3.48\\
$F_{33}$ &  5059.36 &    2.55 &    5.35 &    4.09 &    2.30\\
\hline
&&\multicolumn{4}{c}{$E$}\\
\cline{3-6}
& &3.31&5.78&3.54&3.23\\ 
\hline
\end{tabular}
\end{center}
\end{table}

\begin{figure}
\resizebox{\hsize}{!}{\includegraphics{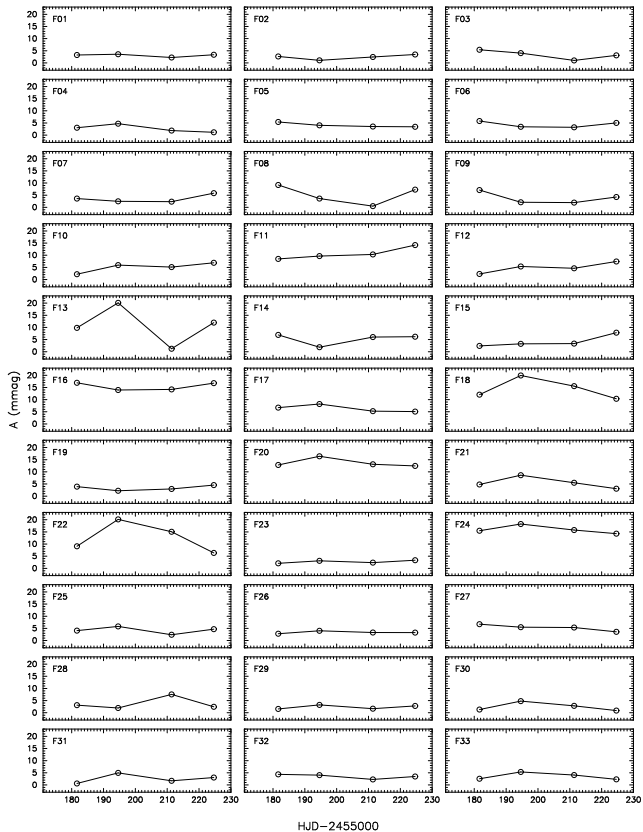}}
\caption{Amplitude variations of the 33 frequencies among four weeks of December of 2009 and January of 2010 in HS~0507+0434B.}
\label{fig6}
\end{figure}

\section{Constraints from the theoretical models}

We computed a grid of DA white dwarf models in a range of the 4 parameters : 0.54 $<$
M/M$_{\odot}$ $<$ 0.75, 2.5$\times 10^{-3}$ $<$ L/L$_{\odot}$ $<$ 6.3$\times 10^{-3}$,
-10$<$$log$q$_{H}$ $<$-4, where $q_{H}$= $M_{H}$/$M_{*}$ is the hydrogen mass fraction and 
for the helium mass fraction  $10^{-4}$ $<$ $M_{He}$/$M_{*}$ $<$ $10^{-2}$. 
In this preliminary modeling, we use the ML2 version of the mixing-length ``theory'' to describe the convection 
with the value of the mixing-length parameter $\alpha$= 0.6, following Bergeron et al. (1995). 
 The models are 
assumed to have reached the equilibrium chemical stratification through the effect of gravitational settling.
Their degenerate core is a homogeneous mixture of carbon and oxygen.
 For each model of
the grid, we computed the periods of the adiabatic $g$-mode oscillations for the degree $\ell$=1 and for orders $k$ = 1 to 27.
A $\chi^{2}$ is computed from the comparison of these periods with the observed periods of the $m$ = 0 central components of the 6 triplets
listed in Table 5.

  Figure~7 shows a 2D slice in a -log($M_{H}$/$M_{*}$) vs $M_{*}$/$M_{\odot}$ plane 
with the values  M$_{He}$/M$_{*}$ = 10$^{-2}$, 
L/L$_{\odot}$=3.5$\times 10^{-3}$, which correspond to the values of these two parameters 
for which  we obtain the best fit in
the complete 4D parameter grid. The best fit model is identified as the one with the lower
$\chi^{2}$. For a better visibility  1/$\chi^{2}$ is shown on the figure; the best-fit model has a 1/$\chi^{2}$= 88.
The best model obtained in the grid corresponds to M$_{*}$/M$_{\odot}$ = 0.675,
L/L$_{\odot}$=3.5$\times 10^{-3}$, M$_{H}$/M$_{*}$=
10$^{-8.5}$. 
 Its effective temperature, 12460~K, is in good agreement with the new spectroscopic estimate of Gianninas et al. (2011) of 12290~$\pm 186$ K. 
The inferred thin hydrogen layer is in agreement with the mode trapping discussed above in section 4.5. 
This low value of the hydrogen mass fraction suggests that HS~0507+0434B has evolved from a last He thermal pulse during which the H is efficiently
burned (Althaus et al. 2005). 

\begin{figure}
\resizebox{\hsize}{!}{\includegraphics{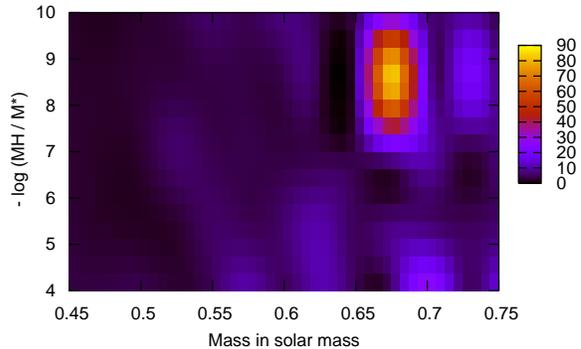}}
\caption{Least square fit of oscillations of HS0507+0434B compared 
with models: 2D cut in the -log($M_{H}$/$M_{*}$) versus $M_{*}$/$M_{\odot}$ of the complete 4D parameters domain. The contour maps show the iso-inverse $\chi^{2}$
of the fits. The lowest value of the contours is for 1/$\chi^{2}$ equals to 5. The next contours are shown from 10 to 80 by step of 10. 
The best fit model corresponds to the value of 1/$\chi^{2}$ equals to 88. }
\label{fig7}
\end{figure}

\begin{table}
\footnotesize
\tabcolsep=1.2mm
\begin{center} \caption{Fit of the computed periods with the  periods observed in HS~0507+0434B in 5 different stellar models; the first column is the model number;
 the following three columns give the total mass, the luminosity and the hydrogen mass fraction of each model; 
the last column gives the inverse $\chi^{2}$ 
of the fit. Model 4 is selected as the best fit model. }
\begin{tabular}{rrrrr}
\hline
\multicolumn{1}{c}{Model } & \multicolumn{1}{c}{$M_{*}/M_{\odot}$} & \multicolumn{1}{c}{$L_{*}/L_{\odot}$($10^{-3}$)} & \multicolumn{1}{c}{$log(M_{H}/M_{*}$)} & \multicolumn{1}{c|}{1/$\chi^{2}$} \\
\hline
1 & 0.700     & 3.2  & -4.0  & 25 \\
2 & 0.675     & 3.3  & -4.0  & 23 \\
3 & 0.675     & 3.4  & -8.5  & 44 \\
4 & 0.675     & 3.5  & -8.5  & 88 \\
5 & 0.675     & 3.6  & -8.5  & 42 \\
\hline
\end{tabular}
\end{center}
\end{table}

\begin{table}
\footnotesize
\tabcolsep=1.2mm
\begin{center} \caption{Preliminary identification of the modes observed in HS~0507+0434B. $P_{obs}$ are the observed periods in seconds, $P_{model}$ the closest
 periods of the best fit model with their degree $\ell$, order $k$ and azimutal number $m$.}
\begin{tabular}{rrrrr}
\hline
\multicolumn{1}{c}{$P_{obs}$} & \multicolumn{1}{c}{$P_{model}$} & \multicolumn{1}{c}{$\ell$} & \multicolumn{1}{c}{$k$} & \multicolumn{1}{c|}{$m$} \\
\hline
197.7 & 198.8 & 2 & 2 &  ? \\
354.9 &       &   &   & +1 \\
355.3 & 357.9 & 1 & 3 &  0 \\
355.8 &       &   &   & -1 \\
&&&&                       \\
444.7 &       &   &   & +1 \\
445.3 & 449.2 & 1 & 5 &  0 \\
446.1 &       &   &   & -1 \\
&&&&                       \\
555.3 &       &   &   & +1 \\
556.5 & 560.5 & 1 & 7 &  0 \\
557.8 &       &   &   & -1 \\
&&&&                       \\
654.8 &       &   &   & +1 \\
655.9 & 654.6 & 1 & 8 &  0 \\
657.1 &       &   &   & -1 \\
&&&&                       \\
695.6 &       &   &   & +1 \\
697.6 & 689.9 & 1 & 9 &  0 \\
699.6 &       &   &   & -1 \\
&&&&                       \\
703.9 & 703.3 & 2 & 18& ?  \\
&&&&                       \\
746.1 &       &   &   & +1 \\
748.6 & 746.0 & 1 & 10&  0 \\
750.3 &       &   &   & -1 \\
&&&&                       \\
972.2 & 972.8 & 2 & 26&  ? \\
999.7 & 986.9 & 2 & 27&  ? \\
1382.7& 1383.0& 2 & 38&  ? \\
\hline
\end{tabular}
\end{center}
\end{table}

Our results are in good agreement with the analysis of Romero et al. (2012) for the total mass and the luminosity (M$_{*}$/M$_{\odot}$= 0.660
and L/L$_{\odot}$= 2.95$\times$ $10^{-3}$) but disagrees on the hydrogen mass fraction for which Romero et al. (2012) find  M$_{H}$/M$_{*}$= 5.7$\times 10^{-5}$.
Note that in their analysis, Romero et al. used the periods of the 4 triplets listed in Castanheira \& Kepler (2009), taken from the study of Handler et al. (2002).
In the present paper we use two more triplets. Note also that our present result for the hydrogen mass fraction is in disagreement with our previous preliminary 
analysis (Fu et al. 2010) in which we also found
a best fit for a ``thick'' hydrogen layer (M$_{H}$/M$_{*}$ = 10$^{-4.4}$). However, in this previous study we introduced the 301~s period in the fit while we are now 
confident 
that this
period results from a linear combination and is not a genuine mode. In addition, the $\chi^{2}$ analysis in the 4D parameter space exhibits many minima; our present fits
 also show a minimum at the previous value of  M$_{H}$/M$_{*}$ = 10$^{-4.4}$ but with a much higher $\chi^{2}$ value than the one we now find for  M$_{H}$/M$_{*}$=
10$^{-8.5}$. The value derived in Fu et al. (2010) comes most probably from the identification of a secondary minimum of the  $\chi^{2}$.

 In estimating the inverse $\chi^{2}$ values for the series of model of our grid, we do find maxima  which sometimes correspond to ``thick'' hydrogen layer models.
 Table 7 summarizes the parameters of the five models which give the best fits among the models computed in our grid. 
f Figure 8 shows the residuals to the average period spacing of the periods computed in each of the five models, from model 1 in the top panel 
to model 5 in the bottom panel. 
All the $\ell$=1 $g$-mode periods of the models are shown in the range of periods corresponding to the observed ones, i.e. between 300~s and 750~s. 
They correspond to radial order $k$ from 4 to 14 for the ``thick'' hydrogen layer models 1 and 2 and $k$ from 2 to 10 for the ``thin'' hydrogen layer models 3, 4 and 5. 
The dotted lines indicate 
the periods of the central component of the six observed triplets.

The observations indicate that the 445~s and the 698~s periods correspond to trapped or ``almost trapped'' modes. In the ``thick'' hydrogen layer model 1, 
these periods would best correspond to the modes of order $k$= 7 and 13, respectively, which are not trapped modes. In model 2, the period fit is the worst and 
the mode ($k$=7) with the closest period  to the observed 445~s does not correspond to a trapped mode. 

In each of the ``thin'' hydrogen layer models 3, 4 and 5, the computed most trapped mode is found for the order $k$=6, with a period of 500.7~s, 491.7~s and 489.8~s, 
respectively. This mode is not detected in our present data sets. As mentioned in $\S 4.5$, the amplitude of a mode is not simply related to its linear growth-rate.
So the amplitude of the trapped mode could have been lower than the detection limit during our observing runs. 
The observed period at 445~s corresponds to the order $k$=5 next to the trapped mode.
The period 698~s corresponds to the order $k$=9. But in model 3 this is not a trapped mode. 
The observed trapped or ``almost-trapped'' modes are in better agreement with the modes computed in model 4 and 5 and the global fit estimated from our inverse 
  $\chi^{2}$ selects the model 4 as the best-fit model.

This discussion illustrates how sensitive the result is on the number of observed and identified modes. 
 In the data sets presented here, the triplet at 698~s seen in the 2007 data is not present in the 2009-2010 data. In addition, the trapped mode around 500~s 
that is predicted in our best fit model is absent in both data sets. 
This justifies the need for a follow up of 
this ZZ Ceti star in order to
 find more  modes to better constrain the model and conclude unambiguously about the ``thick'' versus ``thin'' hydrogen content of that star.
However, there are also other arguments in favor of HS~0507+0434B having a ``thin'' hydrogen layer.   
HS~0507+0434A+B form a common proper motion system. Both their similar distance, 49 pc and 48 pc respectively (Gianninas et al. 2011) and their similar radial velocity,
 within their uncertainties (Maxted et al. 2000, their Table 9), confirm that the two stars have been formed together at the same time. However, 
in spite of their similar  mass 
(M$_{*}$/M$_{\odot}$=0.67 for the A component according to Gianninas et al. and M$_{*}$/M$_{\odot}$=0.675 for the B component 
from our asteroseismic determination), they show significantly 
different effective temperature: the component A having 21550 K while the component B, within the same time scale, has cooled to 12290 K (Gianninas et al. 2011). 
This suggests that HS~0507+0434B should have a thinner hydrogen layer than its companion  HS~0507+0434A.

Coming back to the periods at 301.3~s, 247.4~s and 197.7s which we find to fit well the period distribution with the average period spacing of 49.63~s, as discussed in
$\S 4.2$, we checked that neither of the 247.4~s and 197.7s periods could correspond to genuine $\ell$=1 modes since, at least in our claimed ``best-fit model'',
 the fundamental
$\ell$ =1; $k$ = 1 mode has a period of 271.7~s. 
 However, we will discuss below the case of the 197.7s as a possible $\ell$=2 mode.
 As far as the 301.3~s is concerned, the corresponding frequency (3318~$\mu$Hz) is distant by 150 ~$\mu$Hz from the closest mode
in the best-fit model. As this point, we conclude that our check with the best-fit model confirms these three peaks in the FT as linear combinations.

 Finally, having selected one model as the best fit model, we checked a posteriori whether some of the further signals or linear combinations 
listed in Table 4 could be identified with either 
$\ell$=1 and/or $\ell$=2 modes. To do this we computed the periods of the $\ell$=2 modes of the best fit model for radial orders $k$ from 2 to 49. 
We find that some of the further signals could correspond to $\ell$=2 modes: this is the case for the 703.9~s period ($f_{21}$), the 972.2~s period ($f_{19}$)
and the 999.7~s period ($f_{1}$) . The 1382.7~s period identified as a linear combination ($f_{13}$ - $f_{23}$) in Table 4 
could as well be identified as a $\ell$=2 mode. The 197.7~s identified as another linear combination ($f_{13}$ - $f_{18}$) 
could also be identified as the $\ell$=2, $k$=2 mode for which we find a period of 198.8~s.
 Table 8 summarizes the identifications that can be proposed for as many modes as possible. 
 The observed modes correspond to rather low order $k$, between 3 and 10 for the $\ell$=1 modes. It means that the HS~0507+0434B pulsation spectrum is not 
entirely in the asymptotic regime. The period distribution does not follow exactly the regular period spacing expected in the asymptotic regime. 
This explains that while the ``trapped modes'' are separated by $\Delta$k= 5 in Table 5 it is  $\Delta$k=4 in Table 8.

\begin{figure}
\resizebox{\hsize}{!}{\includegraphics{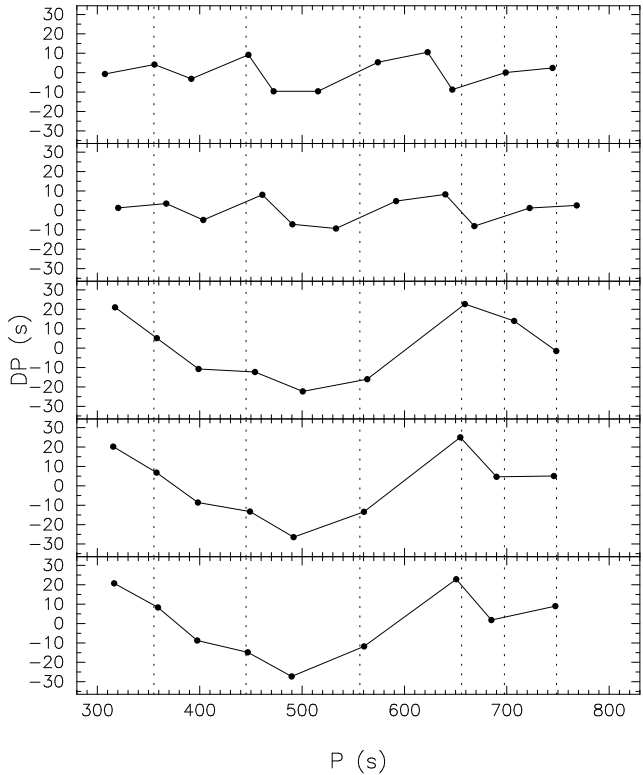}}
\caption{Residuals of the period distribution to the average period spacing for the five best fit models listed in Table 7, from model 1 in the top panel
to model 5 in the bottom panel. 
Dotted lines are the observed periods of the $m$=0 modes of the six identified triplets. Model 4 gives the best fit in terms of $\chi^{2}$. }
\label{fig8}
\end{figure}

\section{Summary and conclusions}

We summarize our results as follows:

 - we have obtained new photometric time-series of the ZZ Ceti star HS~0507+0434B in 2007, 2009 and 2010. The combined power spectra allow us to identify 18 
independent pulsation modes consisting of 6 triplets and one single mode, plus a number of their linear
combinations.  We identify those triplets as $\ell$=1 $g$-modes split by rotation.
  
 - we determine an average period spacing of 49.63~s. The period distribution exhibits the signature of mode trapping.

 - from the frequency shifts measured in the triplets, we determine an average  rotational splitting of 
3.59~$\pm$0.57~$\mu$Hz from which we infer an average rotation rate of 1.61~$\pm$0.26~days. 
 The amplitude ratio of the triplets components,
if interpreted 
 as uniquely due to geometrical effect, could be used to estimate that the angle of the rotation axis 
on the line of sight is close to $70\,^{\circ}$.
 However, the evidence of amplitude variations casts some doubts on the assumption that only geometrical aspect is responsible for the amplitude ratios.

 - the amplitude of the modes vary on week time-scale and we find that the ``pulsation power'' is also time dependent.

 - we computed a grid of models and their $\ell$=1 $g$-modes in the adiabatic approximation. The comparison of their theoretical periods with the observed periods 
through $\chi^{2}$ tests allows us 
to estimate the fundamental parameters of a preliminary  ``best fit'' model which has a total mass of 
 M$_{*}$/M$_{\odot}$ = 0.675, a luminosity
L/L$_{\odot}$=3.5$\times 10^{-3}$, a thin hydrogen outer layer of M$_{H}$/M$_{*}$=
10$^{-8.5}$. 
 The periods of other models with  ``thick'' hydrogen envelope do not fit as well the observed periods according to their $\chi^{2}$.
The ``thin'' hydrogen envelope of the best fit model of HS~0507+0434B is in agreement with the fact that its common proper motion companion HS~0507+0434A,
 which must have been formed at the same time and has a similar mass, has a significantly higher effective temperature.
This implies that the B component must have a thinner H envelope than the A component for having cooled down to a lower effective temperature 
during the same cooling time scale.
This low value of the hydrogen mass fraction suggests that HS~0507+0434B has  evolved from a last He thermal pulse episode.

Further observations of HS~0507+0434B are required to refine the modeling, to better constrain the amplitude variation time-scales and identify the 
physical mechanism driving those variations, and to contribute to the mapping of the convection efficiency through the ZZ Ceti instability strip.

\section*{Acknowledgements}

We acknowledge an anonymous referee for his(her) constructive comments which helped improving a first version of this paper. 
JNF acknowledges the support from the National Natural Science Foundation of
China (NSFC), through the Grants 10878007 and U1231202. The research is partially supported by National Basic Research Program of China (973 Program 2013CB834900) and the Fundamental Research Funds for the Central Universities. ND, GV and SC acknowledge the support from the Programme National de Physique Stellaire (CNRS, INSU). LFM acknowledges financial support from the UNAM under grant PAPIIT IN104612 and from CONACyT by way of grant CC-118611. Special thanks are given to the technical staff and night assistants of 
the Xinglong station of National Astronomical Observatories,
the Lijiang station of Yunnan Astronomical Observatory,
the San Pedro M\'artir Observatory,
the Bohyunsan Optical Astronomy Observatory,
and the Piszk\'estet\H{o} Observatory.
This research has made use of the Simbad database, operated at CDS,
Strasbourg, France.

\end{document}